\documentclass[twocolumn,showpacs,aps,prb,preprintnumbers,amsmath,amssymb]{revtex4}

\usepackage{graphicx}
\usepackage{dcolumn}
\usepackage{bm}
\usepackage{epsfig}

\begin{document}
\date{\today}

\title{
Weak localization and Berry flux in topological crystalline insulators \\ with a quadratic surface spectrum       
}
\author{G. Tkachov }
\affiliation{
Institute for Theoretical Physics and Astrophysics, Wuerzburg University, Am Hubland, 97074 Wuerzburg, Germany}


\begin{abstract} 
The paper examines weak localization (WL) of surface states with a quadratic band crossing 
in topological crystalline insulators.
It is shown that the topology of the quadratic band crossing point dictates the negative sign of the WL conductivity correction. 
For the surface states with broken time-reversal symmetry, an explicit dependence of the WL conductivity on the band Berry flux is obtained and analyzed 
for different carrier-density regimes and types of the band structure (normal or inverted). 
These results suggest a way to detect the band Berry flux through WL measurements.  
\end{abstract}
\maketitle

\section{Introduction}
\label{intro}

Topological insulators (TIs) feature edge or surface states with a gapless spectrum at band crossing points in the Brillouin zone. 
These singularities of the band dispersion have a vortex-like structure and carry quantized Berry's flux 
that contributes to the phase of the electronic wave function, affecting scattering and transport processes.
The best studied example is a {\em linear} (Dirac) band crossing in two-dimensional~\cite{Kane05,Bernevig06,Koenig07} 
and three-dimensional~\cite{Fu07,Murakami07,Moore07} TIs with strong spin-orbit coupling (SOC) (see, also, reviews \onlinecite{Koenig08,Hasan10,Qi10}).
The linear crossing point is protected by time-reversal symmetry (TRS) and carries the Berry flux of $\pi$.  
In this case, the pairs of states with opposite momentum directions appear to be orthogonal to each other and, hence, unavailable for scattering. 
The absence of such backscattering is the hallmark of electron transport in the SOC TI materials 
(see, e.g., reviews \onlinecite{Culcer12,GT13,Ando13}). 
In particular, Dirac surface states escape being localized by potential disorder. Instead, 
the surface conductivity acquires a positive quantum correction, an effect known as weak antilocalization (WAL). 

Recently, a new subclass of TIs - topological crystalline insulators (TCIs) - has been identified.~\cite{Fu11,Hsieh12,Tanaka12,Dziawa12} 
Unlike their SOC counterparts, in the TCIs the gapless surface states are protected by discrete symmetries of the crystal, 
which offers diverse possibilities for engineering and controlling topological states of matter.~\cite{Hsieh12,Tanaka12,Dziawa12}     
A vivid example of the distinct topological properties of the TCIs is 
the possibility of gapless surface states with a {\em quadratic} band crossing.~\cite{Fu11}
These have been predicted for crystalline materials with the fourfold ($C_4$) or sixfold ($C_6$) rotational symmetry on the surface.
The quadratic band degeneracy point is characterized by the Berry flux of $2\pi$, 
which does not forbid backscattering, but nevertheless has implications for quantum transport.~\cite{Novoselov06,Kechedzhi07} 
Most important, instead of WAL the carriers on high-symmetry TCI surfaces are expected to show weak localization (WL), 
with a negative quantum conductivity correction. In contrast to the SOC materials, 
the WL properties of the TCIs still remain unexplored.

In this paper, the WL conductivity correction for the surface states with the quadratic band dispersion 
is calculated by means of Kubo's formalism. Special emphasis is placed on establishing an explicit relation between Berry's flux, $\beta$, 
and the WL conductivity correction, $\delta\sigma$. 
It is shown that $\delta\sigma$ is negative, which is determined by the topology of the quadratic band crossing point. 
If TRS is preserved, there is no other dependence on the band structure, so that the WL correction is typical   
of the orthogonal symmetry class of disordered systems. Richer WL properties are found for the TCIs with broken 
TRS in which the Berry flux can be tuned between $0$ and $4\pi$. 
In this case, the WL shows a unitary behaviour with three characteristic regimes in which the WL conductivity is given per spin by
\begin{eqnarray}
\delta\sigma(\beta) = \frac{e^2}{2\pi h} \times
\left\{
\begin{array}{cc}
 \ln \left[ \frac{\tau_{_0}  }{2\tau_\phi} + \left(\frac{\beta}{4\pi}\right)^2 \right], & \beta \to 0, \\
\\
 \ln \left[ \frac{\tau_{_0} }{\tau_\phi} + 2\left(1 - \frac{\beta}{2\pi}\right)^2 \right], & \beta \to 2\pi,\\
\\
\ln \left[ \frac{\tau_{_0} }{2\tau_\phi} + \left(1-\frac{\beta}{4\pi} \right)^2 \right], & \beta \to 4\pi, 
\end{array}
\right.
\label{dS_beta}
\end{eqnarray}
where $\tau_\phi$ is the dephasing time, and $\tau_{_0} \ll \tau_\phi$ is the characteristic impurity scattering time. 
As explained below, the three cases in Eq. (\ref{dS_beta}) are realized depending on the filling of the conduction band and the type of the band structure (normal or inverted). 
In each case, Eq. (\ref{dS_beta}) establishes a direct link between the intrinsic band Berry flux, $\beta$, and the experimentally accessible observable, $\delta\sigma$. 
This is a distinctly different dependence compared to that found in other TI materials, e.g. 
in magnetically doped three-dimensional TIs,~\cite{He11,Lu11} HgTe quantum wells,~\cite{GT11,Krueckl12,GT13,Muelhlbauer13} 
and doped Kane-Mele TIs.~\cite{Imura09}
   
The subsequent sections provide a comprehensive account of the theoretical approach adopted in this paper. 
In Sec. II a model for the TCI surface state is introduced and incorporated into the general Kubo formalism. 
Section III contains the details of the calculation of the WL conductivity correction based on the solution of the Cooperon equation. 
In Sec. IV the results are summarized and discussed. 

\section{Model}
\label{II}

\begin{figure}[t]
\includegraphics[width=80mm]{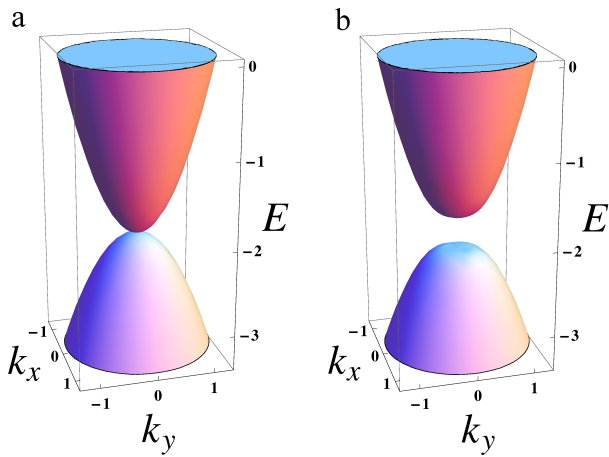}
\caption{(Color online)
Energy band dispersion in the presence (a) and absence (b) of TRS [see, also Eqs. (\ref{H0}) and (\ref{H})].
Fermi level lies in the conduction band at $E=0$. 
}
\label{E_fig}
\end{figure}

\subsection{Effective Hamiltonian and Berry flux}

We consider a 2D system of spinless fermions described by the Hamiltonian
\begin{equation}
\hat{H}_0 = d_1({\bm k})\sigma_z + d_2({\bm k})\sigma_x + d_0({\bm k})\sigma_0,
\label{H0}
\end{equation}
with
\begin{equation}
d_1({\bm k}) = A (k^2_x - k^2_y),\,\,\,
d_2({\bm k}) = B k_x k_y,\,\,\,
d_0({\bm k}) = C {\bm k}^2  - \mu, 
\label{E120}
\end{equation}
where ${\bm k}= (k_x, k_y,0)$ is the wave vector, 
$A$, $B$, and $C$ are band structure constants, and $\mu$ is the Fermi energy. 
The model applies, in particular, to a surface state in crystalline materials of the tetragonal system 
with a diatomic unit cell along the c axis.~\cite{Fu11} In this case, the surface state occurs
on the high symmetry crystal face (001) possessing the fourfold ($C_4$) rotational symmetry. 
Hence, the Pauli matrices $\sigma_x$ and $\sigma_z$ represent the unit-cell degree of freedom, and $\sigma_0$ is the unit matrix. 
In the following we focus on the isotropic case with $B=2A$. 
Hamiltonian (\ref{H0}) is invariant under the time reversal (represented by complex conjugation), 
yielding a gapless spectrum with a quadratic band degeneracy at the high symmetry point ${\bm k}=0$ [see, also Fig. \ref{E_fig}(a)].

We extend the model by adding a TRS-breaking perturbation $\Delta \sigma_y$,  
\begin{equation}
\hat{H} =  \hat{H}_0 + \Delta \sigma_y,
\label{H}
\end{equation}
which opens a gap of $2 |\Delta|$ between the conduction and valence bands at ${\bm k}=0$ [see, also Fig. \ref{E_fig}(b)]. 
This symmetry-breaking mechanism can be incorporated into a spinful model and may result
from a magnetic proximity effect.~\cite{Vobornik11}
The analogy with the magnetic polarization becomes even more pertinent
if one makes a unitary transformation $\hat{U} \hat{H} \hat{U}^\dagger \to {\cal\hat H}$ with matrix
\begin{equation}   
\hat{U} = \frac{i\sigma_0 + \sigma_x + \sigma_y + \sigma_z}{2}, 
\label{U}
\end{equation}
to cast the Hamiltonian in the form 
\begin{equation}
{\cal\hat H} = {\bm d}({\bm k}) \cdot {\bm \sigma} + d_0({\bm k})\sigma_0,
\label{H_c}
\end{equation}
where ${\bm d}({\bm k})$ is the three-component vector  
\begin{figure}[t]
\includegraphics[width=67mm]{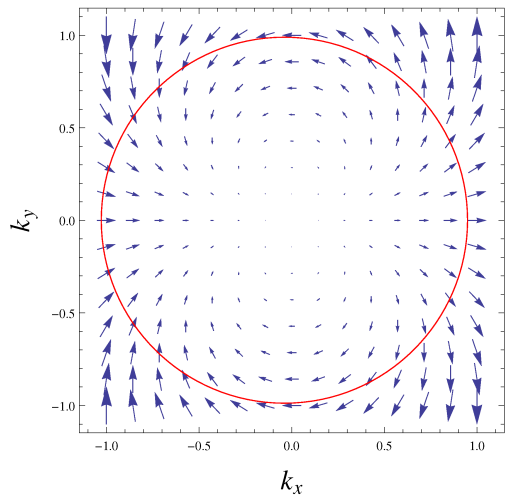}
\caption{(Color online)
Vorticity of vector ${\bm d}({\bm k})$ at ${\bm k}=0$ point [see, also Eq. (\ref{d_vec})]. 
The circle schematically indicates the Fermi surface. All momenta are in units of $A^{-1/2}$. 
}
\label{d_fig}
\end{figure}
\begin{equation}
 {\bm d}({\bm k}) = ( d_1({\bm k}), d_2({\bm k}), \Delta).
\label{d_vec}
\end{equation}
Its out-of-plane component, $\Delta$, accounts for the broken TRS, while the in-plane vector, $( d_1({\bm k}), d_2({\bm k}), 0)$, characterizes 
vorticity associated with the quadratic Fermi point in momentum space (see, Fig. \ref{d_fig}). The vortex carries the Berry flux:
\begin{equation}
\beta = -i \oint_{ |{\bm k}| = k_F } \langle \psi( {\bm k} ) | \nabla_{\bm k} \psi( {\bm k} ) \rangle \cdot d{\bm k}, 
\label{beta_def}
\end{equation}
where the integration path is chosen along a closed Fermi line of radius $k_F$ 
at the crossing of the conduction band with the Fermi level [see, also Figs. \ref{E_fig} and \ref{d_fig}], 
and $\psi({\bm k})$ is the conduction band eigenstate of (\ref{H_c}) given by 
\begin{equation}
\psi({\bm k}) = \frac{ \sigma_0 + {\bm \sigma} \cdot {\bf e}({\bm n}) }{ \sqrt{2(1+{\rm e}_z)} }
\left[
\begin{array}{c}
 1 \\ 0
\end{array}
\right] =
\frac{1}{ \sqrt{2} }
\left[
\begin{array}{c}
 \sqrt{1 + {\rm e}_z} \\ e^{2i\phi} \sqrt{1 - {\rm e}_z}
\end{array}
\right]. 
\label{psi}
\end{equation}
Here ${\bf e}({\bm n}) = {\bm d}/|{\bm d}| = [{\rm e}_x({\bm n}), {\rm e}_x({\bm n}), {\rm e}_z]$ is a unit vector   
describing the vortex structure on the Fermi surface as a function of the unit wave vector, ${\bm n}=[\cos\phi, \sin\phi, 0]$:
\begin{eqnarray}
&&
{\rm e}_x({\bm n}) = {\rm e}_{_\|} (n^2_x - n^2_y) = {\rm e}_{_\|}\cos(2\phi), 
\label{e_x}\\
&& 
{\rm e}_y({\bm n}) = {\rm e}_{_\|} 2n_xn_y = {\rm e}_{_\|}\sin(2\phi),
\label{e_y}\\
&&
{\rm e}_z = \frac{\Delta}{\sqrt{ A^2k^4_F + \Delta^2 } }, \quad {\rm e}_{_\|} = \sqrt{1-{\rm e}^2_z}
\label{e_z}
\end{eqnarray}
In view of the $\pi$-periodicity of the vortex structure and broken TRS, the Berry flux is 
\begin{equation}
\beta = 2\pi ( 1 - {\rm e}_z ) = 2\pi \left( 1 - \Delta/\sqrt{ A^2k^4_F + \Delta^2 }  \right).
\label{beta}
\end{equation}
We assume a simple relation, $k_F = \sqrt{4\pi n}$, between the Fermi wave number and surface carrier density, $n$. 
Table~\ref{table} shows the values of $\beta$ close to modulo $2\pi$ depending on the type of the band structure and the carrier-density regime. 
The characteristic carrier density, $n_0$, is given by 
\begin{equation}
n_0 = |\Delta|/4\pi A.
 \label{n0}
\end{equation}
For high carrier densities,  $ n \gg n_0$ (or $\mu \gg \Delta$), the Berry flux is close to $2\pi$ 
independently of the band structure type. 
For low carrier densities, $n \ll n_0$ (or $\mu \to |\Delta|$), the behaviour of $\beta$ depends on whether the band structure is normal ($\Delta >0$) or inverted ($\Delta < 0$). 
For the normal structure $\beta \to 0$, while for the inverted one $\beta \to 4\pi$. 
Other examples of materials with nontrivial quadratic band dispersion and Berry's phases include 
semiconductor hole structures (see, e.g., Refs. \onlinecite{Jungwirth02,Zhou07,Jaaskelainen10,Krueckl11})
and bilayer graphene (see, e.g., Ref. \onlinecite{Novoselov06}).

\begin{table}[t]
\centering
\begin{tabular}{|l|c|c|c|c|c|c|c|c|}\hline
                              & $n \gg n_0 $       &  $n \ll n_0$   \\ \hline
Normal band structure ($\Delta> 0$)         & $\beta\to 2\pi$                                      &   $ \beta \to 0$                                  \\ \hline
Inverted band structure ($\Delta< 0$)       & $\beta\to 2\pi$                                    &   $\beta \to 4\pi $                               \\ \hline
\end{tabular}
\caption{
Characteristic values of Berry flux, $\beta$, for different types of the band structure and carrier-density regimes. 
}
\label{table}
\end{table}

\subsection{Kubo formula. Model of disorder}

To calculate the electric conductivity, we use the linear response theory with respect to an external uniform electric field, 
${\cal E}e^{-i\omega t}$, at frequency $\omega$. The longitudinal conductivity is given by Kubo formula     
\begin{eqnarray}
\sigma_{xx} &=&  
\frac{e^2}{2\pi \omega a} \int dE [f(E) - f(E+\hbar\omega)]\times
\nonumber\\
&&
\sum_{ {\bm k}, {\bm k^\prime}  } 
{\rm Tr}[ 
\hat{v}^x_{\bm k} \hat{G}^{^R}_{ {\bm k}, {\bm k^\prime} }(E+\hbar\omega) 
\hat{v}^x_{\bm k^\prime} \hat{G}^{^A}_{ {\bm k^\prime}, {\bm k} }(E) 
],  
\label{Kubo}
\end{eqnarray}
where $a$ is the area of the system, $f(E)$ is the Fermi-Dirac distribution function, 
${\rm Tr}$ denotes the trace in ${\bm \sigma}$ space, $\hat{v}^x_{\bm k}$ is the $x$-component of the velocity operator 
\begin{equation}
\hat{v}^x_{\bm k} = \frac{1}{\hbar} \frac{\partial {\cal\hat H} }{ \partial k_x } = 
\frac{2A}{\hbar} {\bm k} \cdot {\bm \sigma} + \frac{2C}{\hbar}k_x\sigma_0,
\label{v}
\end{equation}
and $\hat{G}^{^{R/A}}_{ {\bm k}, {\bm k^\prime} }(E)$ are the retarded and advanced Green functions satisfying the equation 
\begin{equation} 
\hat{G}^{^{R/A}}_{ {\bm k}, {\bm k}^\prime } = \hat{G}^{^{R/A}}_{0 {\bm k} } \delta_{ {\bm k}, {\bm k}^\prime } 
+ \sum_{ {\bm k}_1 } \hat{G}^{^{R/A}}_{0 {\bm k} } \hat{V}_{ {\bm k}, {\bm k}_1} \hat{G}^{^{R/A}}_{ {\bm k}_1, {\bm k^\prime} }.
\label{ExactI_G}
\end{equation}
In the above equations and throughout the "hat" indicates $2\times 2$ matrices in ${\bm \sigma}$ space. 
$\hat{G}^{^{R/A}}_{0 {\bm k} }$ are the bare Green functions defined by the equation 
$
(E - \hat{\cal H} )\hat{G}^{^{R/A}}_{0 {\bm k} } = \sigma_0.
$
Assuming the splitting between the conduction and valence bands at $|{\bm k}|=k_F$ to be much larger 
than the characteristic scale of $E$,
\begin{equation}
 2\sqrt{  A^2 k^4_F + \Delta^2 } = 2( \mu  - C k^2_F ) \gg E,
\label{A_large}
\end{equation}
we find $\hat{G}^{^{R/A}}_{0 {\bm k} }$ near the Fermi surface ($|{\bm k}| \approx k_F$) as 
\begin{equation}
\hat{G}^{^{R/A}}_{0 \bm k} \approx 
\frac{ \hat{P}_{\bm n} }{ E - \xi_{\bm k } }, \qquad
\hat{P}_{\bm n} = \frac{ \sigma_0 + {\bm \sigma} \cdot {\bf e}({\bm n}) }{2},
\label{G0}
\end{equation}
where $\hat{P}_{\bm n}$ is the projector on the conduction band, and
$\xi_{\bm k } = \sqrt{  A^2 k^4 + \Delta^2 } + C k^2 - \mu$ is the conduction band dispersion.

Finally, $\hat{V}_{ {\bm k}, {\bm k}_1}$ in Eq. (\ref{ExactI_G}) is the matrix element of the scattering potential. 
We consider scattering from a spin-independent short-ranged random potential characterized by the correlation function 
\begin{equation}
\langle\langle 
\hat{V}_{ {\bm k}, {\bm k}_1 } \otimes \hat{V}_{ {\bm k}^\prime, {\bm k}_2 }
\rangle\rangle  
= 
\frac{ \zeta }{a}  
\delta_{ {\bm k} - {\bm k}_1, - {\bm k}^\prime + {\bm k}_2}
\sigma_0 \otimes \sigma_0, \quad 
\zeta = \frac{\hbar}{\pi N \tau_{_0}},
\label{Corr_kk}
\end{equation}
where the double brackets $\langle\langle...\rangle\rangle$ denote averaging over the ensemble of the disorder realizations, 
and $\otimes$ indicates the direct matrix product. The correlation strength, $\zeta$, is parametrized 
in terms of the characteristic scattering time, $\tau_{_0}$, and the density of states (DOS) at the Fermi level per spin, $N$. 

\begin{figure}[b]
\includegraphics[width=80mm]{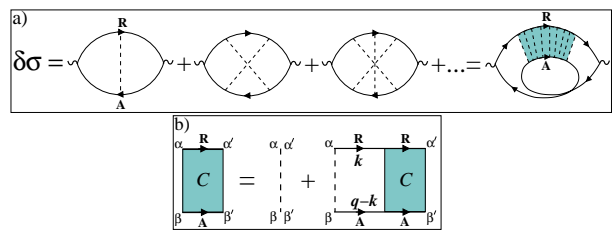}
\caption{(Color online)
Diagrammatic representations for (a) Kubo formula for the quantum correction to Drude conductivity, and (b) 
Bethe-Salpeter equation for the Cooperon. Solid lines with arrows correspond to the disorder-averaged retarded (R) and advanced (A) 
Green functions (\ref{G_RA}), dashed lines to the disorder correlator (\ref{Corr_kk}), and shaded area to the Cooperon. 
}
\label{Diagrams}
\end{figure}

\section{ Theoretical approach }

\subsection{Quantum correction to classical conductivity from Kubo formula}
\label{ss_dS}

We follow the standard approach in which Kubo formula (\ref{Kubo}) is averaged over the disorder configurations, and 
the quantum correction to Drude conductivity, $\delta\sigma$, is given by the crossed diagrams summing up into the Cooperon 
$C_{\alpha\beta\alpha^\prime\beta^\prime}({\bm q}, \omega)$ as depicted in Fig. \ref{Diagrams} 
(see, e.g., Refs. \onlinecite{Altshuler80,Rammer04}). The corresponding analytic expression for $\delta\sigma$  is
\begin{widetext}
\begin{eqnarray}
\delta\sigma =
\frac{e^2\hbar}{\pi N \tau^2}
\int\frac{dE}{2\pi\omega} [ f(E) - f(E+\hbar\omega) ] 
\sum_{\alpha\beta\alpha^\prime\beta^\prime}
&&
\int\frac{ d{\bm k} }{(2\pi)^2}
(
\hat{G}^{^A}_{{\bm k}, E}\,
\hat{v}^x_{\bm k}\,
\hat{G}^{^R}_{ {\bm k}, E+\hbar\omega }
)_{\beta^\prime\alpha}
(
\hat{G}^{^R}_{ {\bm q}-{\bm k}, E+\hbar\omega }\,
\hat{v}^x_{ {\bm q}-{\bm k} }\,
\hat{G}^{^A}_{ {\bm q}-{\bm k}, E }
)_{\alpha^\prime\beta}
\times
\qquad\label{dS}\\
&&
\int\frac{ d{\bm q} }{(2\pi)^2} C_{\alpha\beta\alpha^\prime\beta^\prime}({\bm q},\omega),
\nonumber
\end{eqnarray}
\end{widetext}
where the Greek indices label the states in ${\bm \sigma}$ space. 
The Cooperon obeys the Bethe-Salpeter equation [see, also Fig. \ref{Diagrams}(b)], 
\begin{eqnarray}
&&
C_{\alpha\beta\alpha^\prime\beta^\prime}({\bm q},\omega)= 
\frac{\tau^2}{\tau_{_0}}\delta_{\alpha\alpha^\prime}\delta_{\beta\beta^\prime} 
+ 
\zeta
\int \frac{d{\bm k}}{(2\pi)^2}\times
\label{Eq_C}\\
&&
\sum_{\gamma^\prime\delta^\prime}
G^{^R}_{\alpha\gamma^\prime}({\bm k}, E +\hbar\omega )
G^{^A}_{\beta\delta^\prime}({\bm q} - {\bm k}, E) 
C_{\gamma^\prime\delta^\prime\alpha^\prime\beta^\prime}({\bm q},\omega).
\quad
\nonumber
\end{eqnarray}
Due to the chosen normalization of $C_{\alpha\beta\alpha^\prime\beta^\prime}({\bm q},\omega)$,
the prefactors in Eqs. (\ref{dS}) and (\ref{Eq_C}) contain the elastic scattering time, $\tau$, given by
\begin{equation}
\frac{1}{\tau} = \frac{2\pi N \zeta}{\hbar} 
\int^{2\pi}_{0} \frac{ d\phi_{\bf n_1} }{2\pi} 
\frac{ 1 + {\bf e}({\bm n}) \cdot {\bf e}({\bm n_1}) }{2} = \frac{1 + {\rm e}^2_z}{\tau_{_0}},  
\label{tau}
\end{equation}
where the integration goes over the directions of the scattered state specified by the unit momentum vector 
${\bm n_1}= (\cos\phi_{\bm n_1}, \sin\phi_{\bm n_1}, 0)$. 
The same time $\tau$ enters the disorder-averaged Green functions $\hat{G}^{^{R/A} }_{{\bm k},E}$ in Eqs. (\ref{dS}) and (\ref{Eq_C}), 
\begin{eqnarray}
\hat{G}^{^{R/A} }_{{\bm k},E} =\frac{ \hat{P}_{\bm n} }
{E_{_{R/A} } - \xi_{\bm k } },
\qquad
E_{_{R/A} }= E \pm \frac{ i\hbar}{2\tau}.
\label{G_RA}
\end{eqnarray}
Since velocity operator (\ref{v}) is odd in ${\bm k}$, the vertex corrections vanish identically. 
As a result, the transport scattering time coincides with $\tau$, 
the diffusion constant is $D = v^2\tau/2$, and  
there are no additional corrections to the conductivity in Eq. (\ref{dS}).   
Equations (\ref{dS}) - (\ref{G_RA}) are valid in the metallic regime 
\begin{equation}
k_F v \tau \gg 1.
\label{metal}
\end{equation}

Since under conditions (\ref{A_large}) and (\ref{metal}) only the vicinity of the Fermi surface matters, 
we employ the standard integration over $\xi_{\bm k}$ in Eq. (\ref{dS}), 
after which the conductivity correction assumes the form~\cite{Integrals} 
\begin{eqnarray}
\delta\sigma &=& 
\frac{8e^2D}{hv^2}
\sum_{\alpha\beta\alpha^\prime\beta^\prime}
\overline{ 
[
\hat{P}_{\bm n} {\hat v}^x_{\bm n} \hat{P}_{\bm n}
]_{\beta^\prime\alpha}
[
\hat{P}_{-{\bm n}} {\hat v}^x_{-{\bm n} } \hat{P}_{-{\bm n}}
]_{\alpha^\prime\beta}
}
\nonumber\\
&\times&
\int\frac{d{\bm q}}{(2\pi)^2}C_{\alpha\beta\alpha^\prime\beta^\prime}({\bm q},\omega),
\label{dS_1}
\end{eqnarray}
where the bar denotes averaging over the directions of the unit vector ${\bm n}$: 
$
\overline{ (...) } = \int^{2\pi}_0 ... d\phi_{\bm n}/(2\pi).
$
In order to sum out the spin degrees of freedom, we expand the Cooperon in the orthonormal basis  
of the two-electron spin states,
\begin{eqnarray}
C_{\alpha\beta\alpha^\prime\beta^\prime}({\bm q},\omega)=\sum_{ij}
C^{ij}({\bm q},\omega)\, \Psi^i_{\alpha\beta}\Psi^{j*}_{\alpha^\prime\beta^\prime},
\label{Expansion}
\end{eqnarray}
where the basis functions, $\Psi^i_{\alpha\beta}$, can be chosen as follows

\begin{eqnarray}
 \Psi^j=\frac{\sigma_j\sigma_y}{ \sqrt{2} }, \quad j=0,x,y,z, \quad 
\sum_{\alpha\beta}
\Psi^j_{\alpha\beta}\Psi^{i*}_{\alpha\beta}=\delta_{ji}. 
\label{Psi}
\end{eqnarray}
The index $j=0$ labels the singlet state, while $j= x, y$, and $z$ correspond to the three triplet states. 
After the straightforward summation with the use of Eqs. (\ref{Expansion}) and (\ref{Psi}), we find
\begin{eqnarray}
\delta\sigma &=&
-\frac{4e^2D}{h} \int\frac{ d{\bm q}  }{(2\pi)^2}
\bigl[
\overline{ n^2_x ( 1 - {\rm e}^2_x ) } \, C^{xx} 
+
\overline{ n^2_x (1-{\rm e}^2_y) }  \, C^{yy}
\nonumber\\
&+&
\overline{ n^2_x (1-{\rm e}^2_z) }  \, C^{zz}
+
\overline{ n^2_x {\rm e}_z} \, i(C^{xy} - C^{yx})
\bigr].
\label{dS_2}
\end{eqnarray}
This expression reflects the $\pi$-periodicity of the vortex structure in momentum space, with ${\bf e}(-{\bm n})={\bf e}({\bm n})$. 
Because of that, there is no contribution of the singlet Cooperon, $C^{00}$, which is responsible for the WAL in the Dirac systems.~\cite{Suzuura02}  
The correction (\ref{dS_2}) is always negative.~\footnote{In the TCIs with an even (e.g. two) number of the surface Dirac cones the WAL and WL processes may compete, 
depending on the strength of the scattering between the Dirac points.}
At the same time, Eq. (\ref{dS_2}) differs from the WL conductivity of a conventional 2DEG with a ${\bm \sigma}$-independent 
quadratic Hamiltonian. To illustrate the difference, in Appendix \ref{A} we obtain the WL correction for the conventional 2DEG from Eq. (\ref{dS_1}). 
We note that a negative WL conductivity has also been found for the semiconductor hole systems under appropriate conditions 
(see, e.g., Refs. \onlinecite{Averkiev98}, \onlinecite{Krueckl11}, and \onlinecite{Porubaev13}) and for bilayer graphene.~\cite{Kechedzhi07}     

Apart from the diagonal triplet Cooperons, Eq. (\ref{dS_2}) contains the off-diagonal ones, $C^{xy}$ and $C^{yx}$.
These are induced by the polarization term and both proportional to $i{\rm e}_z$. 
Technically, the off-diagonal Cooperon terms originate from the matrix elements 
$\sigma^{\beta^\prime\alpha}_0\sigma^{\alpha^\prime\beta}_z {\rm e}_z$ and 
$\sigma^{\beta^\prime\alpha}_z\sigma^{\alpha^\prime\beta}_z {\rm e}_z$ in the prefactor in front of the integral in Eq. (\ref{dS_1}).
Without the off-diagonal Cooperons $C^{xy}$ and $C^{yx}$ Eq. (\ref{dS_2}) 
cannot correctly describe the case of strong polarization, $|{\rm e}_z|\to 1$. 
In the next subsection we calculate the required triplet Cooperon amplitudes. 

\subsection{Cooperon amplitudes}
\label{ss_C}

The equation for the Cooperon amplitudes $C^{ij}({\bm q},\omega)=\sum_{\alpha\beta\alpha^\prime\beta^\prime}
\Psi^{i*}_{\alpha\beta} \Psi^{j}_{\alpha^\prime\beta^\prime}\,C_{\alpha\beta\alpha^\prime\beta^\prime}(\omega, {\bm q})$ 
follows from Eq. (\ref{Eq_C}). After the standard integration procedure~\cite{Integrals} we find
\begin{equation}
C^{ij}({\bm q},\omega)= 
\frac{\tau^2}{\tau_{_0}}\delta_{ij} 
+ 
\frac{\tau}{\tau_{_0}} 
\sum_{s=0,x,y,z}
{\rm Tr}
\langle
\hat{P}_{- {\bm n}} 
\sigma_i 
\hat{P}_{\bm n}  
\sigma_s
\rangle
C^{sj}({\bm q},\omega),
\label{Eq_Cij}
\end{equation}
where the brackets $\langle ... \rangle$ stand for the integral
\begin{equation}
\langle...\rangle
=\int_0^{2\pi} \frac{ d\phi_{\bm n} }{2\pi}
\frac{...}{ 1 - i \tau\omega + i\tau v {\bm n}\cdot{\bm q} }, 
\label{Angle}
\end{equation}
taken over the momentum directions on the Fermi surface.
Equation (\ref{Eq_Cij}) reproduces the Cooperon amplitudes for the conventional 2DEG (see, Appendix \ref{A}). 
In our case, the specifics of the system consists in the $\pi$-periodic pseudospin texture determined 
by vector ${\bf e}({\bm n})$ in Eqs. (\ref{e_x}) - (\ref{e_z}). 
First, we make use of the fact that ${\bf e}({\bm n})$ is an even function of ${\bm n}$. 
This allows us to reduce Eq. (\ref{Eq_Cij}) to

\begin{widetext}
\begin{equation}
\left[
\begin{array}{ccc}
1 + {\rm e}^2_z - \langle 1-{\rm e}^2_x  \rangle &
\langle i{\rm e}_z + {\rm e}_x{\rm e}_y \rangle & 
\langle -i{\rm e}_y + {\rm e}_x {\rm e}_z \rangle\\
\langle  -i{\rm e}_z  + {\rm e}_y {\rm e}_x \rangle & 
1 + {\rm e}^2_z - \langle 1-{\rm e}^2_y \rangle &
\langle i{\rm e}_x + {\rm e}_y{\rm e}_z \rangle\\
\langle  i{\rm e}_y + {\rm e}_z{\rm e}_x \rangle & 
\langle -i{\rm e}_x + {\rm e}_z{\rm e}_y \rangle &
1 + {\rm e}^2_z - \langle 1-{\rm e}^2_z \rangle
\end{array}
\right]
\left[
\begin{array}{ccc}
 C^{xx} &  C^{xy} & C^{xz} 
\\ 
C^{yx}  &  C^{yy} & C^{yz}
\\ 
C^{zx} & C^{zy} & C^{zz}
\end{array}
\right]
=
\tau
\left[
\begin{array}{ccc}
 1 & 0 & 0 
\\ 
0  & 1 & 0 
\\ 
0  & 0 & 1
\end{array}
\right], 
\label{Cxyz}
\end{equation}
\end{widetext}
and $ C^{0j}= (\tau^2/\tau_{_0}) \delta_{0j}$. 
Note that the Cooperons with the singlet first index ($i=0$) decouple from the rest and are independent of ${\bf q}$ and $\omega$. 
In Appendix \ref{B} we solve Eq. (\ref{Cxyz}) in the diffusion approximation 
\begin{equation}
\tau v {\bm n} \cdot {\bm q} \ll 1, \qquad \tau\omega \ll 1. 
\label{diff}
\end{equation}
For the required Cooperon amplitudes we find 
\begin{eqnarray}
C^{xx}({\bm q},\omega) = 
&\frac{1}{2}&
\left[
\frac{1}{ (1+{\rm e}_z)( D{\bm q}^2 -i\omega ) + (1-{\rm e}_z)^2/2\tau } 
\right.
\qquad\label{Cxx}\\
&+&
\left.
\frac{1}{ (1-{\rm e}_z)( D{\bm q}^2 -i\omega ) + (1+{\rm e}_z)^2/2\tau }
\right],
\nonumber
\end{eqnarray}

\begin{equation}
C^{yy} = C^{xx}, \qquad C^{yx} = - C^{xy} = \frac{ 2i{\rm e}_z }{ 1 + {\rm e}^2_z } \, C^{xx},
\label{Cxy}
\end{equation}

\begin{equation}
C^{zz}({\bm q},\omega) = \frac{1}{ (1-{\rm e}^2_z)( D{\bm q}^2 -i\omega ) + 2{\rm e}^2_z/\tau }. 
\label{Czz}
\end{equation}
In these equations, parameter ${\rm e}_z$ quantifies the degree to which the TRS is broken by the polarization field.
In the TRS case (${\rm e}_z=0$), Cooperon $C^{zz}$ is gapless, while $C^{xx}$ and $C^{yy}$ both have a large relaxation gap 
of $1/2\tau$. If the TRS is broken, all the Cooperons are gapped as expected for the unitary universality class. 
For a weak polarization with ${\rm e}^2_z\ll 1$ a small gap of $2{\rm e}^2_z/\tau$ opens in $C^{zz}$.
For a strong polarization with $|{\rm e}_z| \to 1$, Cooperons $C^{xx}$ and $C^{yy}$ become gapless,
while the gap in $C^{zz}$ increases to $2/\tau$. In each case, the conductivity correction is determined by 
the Cooperons with a small gap.        

\section{ Results }

The TRS breaking parameter, ${\rm e}_z$, can be controlled by tuning the carrier density, $n$  [see, Eqs. (\ref{e_z}) and (\ref{beta}) and text after].
We begin by evaluating WL conductivity (\ref{dS_2}) at high carrier densities 
\begin{equation}
n \gg n_0,
\label{high_n}
\end{equation}
which corresponds to $|{\rm e}_z| \ll 1$.  In this case, 
\begin{equation}
C^{zz}({\bm q},\omega) \approx \frac{1}{ D{\bm q}^2 -i\omega + 2{\rm e}^2_z/\tau }, 
\label{Czz_1}
\end{equation}
and the other Cooperons can be neglected. 
With the upper integration cutoff $Dq^2_c = \tau^{-1} \approx \tau^{-1}_0$ and replacement $-i\omega \to \tau^{-1}_\phi$, 
Eqs. (\ref{dS_2}) and (\ref{Czz_1}) yield  
\begin{equation}
\delta\sigma \approx \frac{e^2}{2\pi h}\ln\left(  \frac{\tau_{_0}}{\tau_\phi} + 2 {\rm e}^2_z \right).
\label{dS_high}
\end{equation}
The case of strong polarization ($|{\rm e}_z| \to 1$) corresponds to low carrier densities 
\begin{equation}
n \ll n_0.
\label{low_n}
\end{equation}
Under this condition Cooperons $C^{xx/yy}$ (\ref{Cxx}) and $C^{xy/yx}$ (\ref{Cxy}) are  
\begin{eqnarray}
C^{xx} = C^{yy} \approx  
\frac{1/2}{ 2( D{\bm q}^2 -i\omega ) + (1-|{\rm e}_z|)^2/2\tau }, 
\label{Cxx_1}
\end{eqnarray}
\begin{eqnarray}
C^{yx} = - C^{xy} \approx i \, {\rm sgn}({\rm e}_z) \, C^{xx},
\end{eqnarray}
and $C^{zz}$ is negligible. With the upper integration cutoff $Dq^2_c = \tau^{-1} \approx 2 \tau^{-1}_0$ in Eq. (\ref{dS_2}), 
we find 
\begin{equation}
\delta\sigma \approx \frac{e^2}{2\pi h} \ln \left( \frac{\tau_{_0}}{2\tau_\phi} + \frac{ (1 - |{\rm e}_z|)^2}{4} \right).
\label{dS_low}
\end{equation}
In view of Eq. (\ref{beta}) the dependence on the polarization ${\rm e}_z$ is equivalent to the dependence on the Berry flux, $\beta$.
Replacing ${\rm e}_z$ by $\beta$ in Eqs. (\ref{dS_high}) and (\ref{dS_low}), we arrive at Eq. (\ref{dS_beta}) for the WL conductivity 
correction, $\delta\sigma(\beta)$, announced in the introduction. We note that the regimes ${\rm e}_z \to \pm 1$ are realized in the normal 
and inverted structures with $\beta \to 0$ and $\beta\to 4\pi$, respectively. 
 
\begin{figure}[t]
\includegraphics[width=80mm]{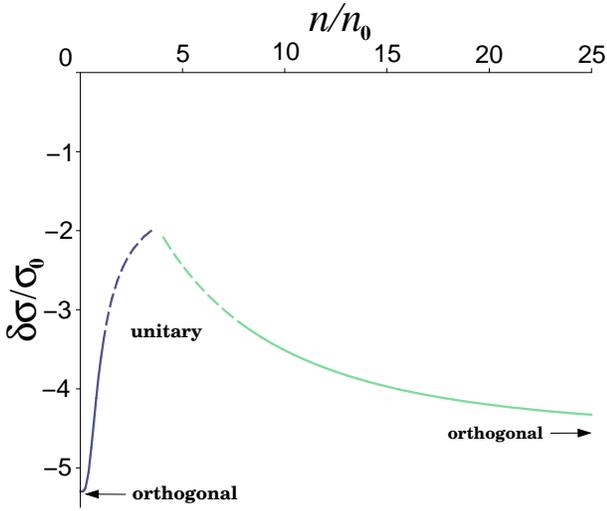}
\caption{(Color online)
WL conductivity correction, $\delta\sigma$, versus carrier density, $n$ [see, also Eqs. (\ref{dS_high}) and (\ref{dS_low})];
$\sigma_0 = e^2/(2\pi h)$ and $\tau_\phi/\tau_{_0} =100$. 
}
\label{dS_fig}
\end{figure}

\begin{figure}[t]
\includegraphics[width=80mm]{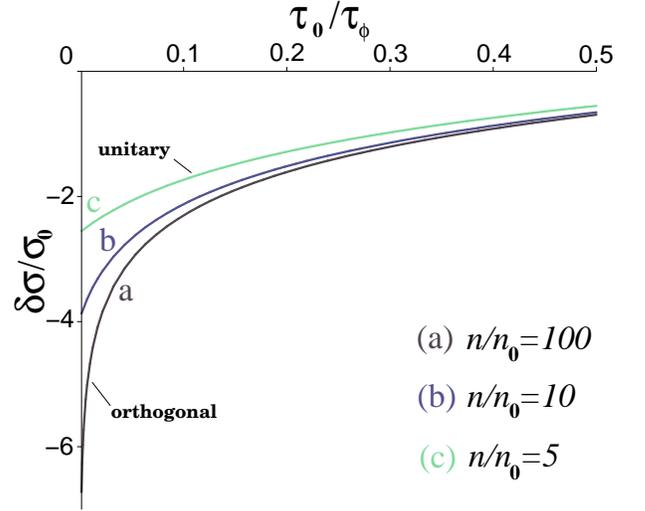}
\caption{(Color online)
WL conductivity correction, $\delta\sigma$, versus normalized dephasing rate, $\tau_{_0}/\tau_\phi$, for different carrier densities 
[see, also Eq. (\ref{dS_high})]. 
}
\label{dS_t_fig}
\end{figure}

In conclusion, we discuss two possible experimental signatures of the Berry-flux dependence of $\delta\sigma$. 
One is a nonmonotonic carrier-density dependence of $\delta\sigma$, which follows from asymptotics (\ref{dS_high}) and (\ref{dS_low}).
In Fig. \ref{dS_fig}  Eqs. (\ref{dS_high}) and (\ref{dS_low}) are plotted as a function of the normalized density, $n/n_0$. 
In the two extreme limits $n/n_0 \to 0$ and $n/n_0 \to \infty$ the system behaviour is typical of the orthogonal universality class with 
a logarithmically large negative $\delta\sigma \approx -(e^2/2\pi h)\ln (\tau_\phi/\tau_{_0})$. 
On the crossover between the orthogonal limits at $n \sim n_0$, the WL correction should reach a maximum value of order of 
$-e^2/2\pi h$. The maximum is the signature of the Berry flux in the well-developed unitary regime in which the phase-coherent quantum interference 
is limited by a short time-scale $\sim \tau$.~\footnote{We note that, unlike the Dirac materials, here the Berry flux of $\pi$ is achieved in a system with a broken TRS (short phase-coherence time) and does not lead to the WAL.}
This corresponds to the large gaps in the Cooperon amplitudes in Eqs. (\ref{Cxx}) - (\ref{Czz}).   
The other possibility is to examine the dependence of $\delta\sigma$ on the dephasing rate $\tau^{-1}_\phi$, 
as shown in Fig. \ref{dS_t_fig}. For a sufficiently large carrier density (curve a in Fig. \ref{dS_t_fig})  
the correction $\delta\sigma$ tends to be divergent in the limit $\tau^{-1}_\phi \to 0$. This indicates that $\beta$ is very close to $2\pi$. 
For somewhat lower densities (e.g., curve c in Fig. \ref{dS_t_fig}) $\delta\sigma$ is less sensitive to $\tau^{-1}_\phi$. 
Its finite value at $\tau^{-1}_\phi\to 0$, 
\begin{equation}
\delta\sigma \approx \frac{e^2}{\pi h} \ln \left| 1- \frac{\beta}{2\pi} \right|,
\label{dS_0} 
\end{equation}
is the measure of the Berry flux, $\beta$. 
Experimentally, this limit can be achieved at sufficiently low temperatures. 

\appendix
\section{WL correction and Cooperon amplitudes for a conventional 2DEG from Eqs. (\ref{dS_1}) and (\ref{Eq_Cij})}
\label{A}

Since matrix $\hat{P}_{\bm n}$ in Eq. (\ref{dS_1}) is still arbitrary,  
this equation is valid also for a 2DEG with a spin-independent quadratic Hamiltonian.
In this case, $A=B=0$ in Eqs. (\ref{H0}) and (\ref{v}), and the Green functions are given by Eq. (\ref{G_RA}) 
where $\hat{P}_{\bm n}$ should be replaced by the unit matrix, $\sigma_0$. 
Then, using Eq. (\ref{Expansion}) we find from Eq. (\ref{dS_1})
\begin{eqnarray}
\delta\sigma &=& 
- \frac{4e^2D}{h} \int \frac{d{\bm q}}{(2\pi)^2} \sum_{\alpha\beta} C_{\alpha\beta\beta\alpha}
\label{dS_metal}\\
&=& 
- \frac{4e^2D}{h} \int \frac{d{\bm q}}{(2\pi)^2} \sum_{ij} C^{ij}{\rm Tr}[\Psi^i\Psi^{j*}] 
\nonumber\\
&=& 
- \frac{4e^2D}{h} \int \frac{d{\bm q}}{(2\pi)^2} [ C^{xx} + C^{yy} + C^{zz} - C^{00} ].  
\nonumber
\end{eqnarray}
Unlike Eq. (\ref{dS_2}), the above equation involves both triplet and singlet diagonal Cooperons.   
These can be calculated from Eq. (\ref{Eq_Cij}) with $\hat{P}_{\bm n}=\sigma_0$ and $\tau = \tau_{_0}/2$ as 
\begin{equation}
C^{ij}({\bm q},\omega)= \frac{\tau/2}{ 1 - \langle 1 \rangle } \delta_{ij},
\label{C^ij}
\end{equation}
where $\langle 1 \rangle$ is the angle integral
\begin{equation}
\langle 1 \rangle
=
\int_0^{2\pi} \frac{ d\phi_{\bm n} }{2\pi}
\frac{1}
{1 - i \tau\omega + i\tau v {\bm n}\cdot{\bm q} } \approx 1 + i\tau\omega - \frac{\tau^2 v^2 {\bm q}^2}{2}, 
\label{Angle_1}
\end{equation}
under conditions (\ref{diff}). Inserting this into Eq. (\ref{C^ij}), we have 
\begin{equation}
C^{ij}({\bm q},\omega)= \frac{1/2}{ D{\bm q}^2 - i\omega  } \delta_{ij}, \qquad D = \frac{v^2\tau}{2}. 
\label{C^ij_1}
\end{equation}
The factor of 1/2 is due to the chosen normalization of the Cooperon. 
Equations (\ref{dS_metal}) and (\ref{C^ij_1}) lead to the well known result: 
$\delta\sigma = -\frac{2e^2}{(2\pi)h}\ln \frac{\tau_\phi}{\tau}$, 
where the factor of 2 in the numerator accounts for the band degeneracy.  

\section{Cooperon amplitudes from Eq. (\ref{Cxyz})}\label{B}

We seek the solutions of Eq. (\ref{Cxyz}) with the diffusion pole structure similar to that in Eq. (\ref{C^ij_1}).
Let us first estimate the off-diagonal matrix elements $\langle {\rm e}_{x,y} \rangle$ and 
$\langle {\rm e}_x{\rm e}_y \rangle$ in Eq. (\ref{Cxyz}).
To do so we expand the denominator in Eq. (\ref{Angle}) and perform averaging over ${\bm n}$ under conditions (\ref{diff}) .
Since ${\rm e}_{x,y}$ are both second harmonics of $\phi$ [see, Eqs. (\ref{e_x}) and (\ref{e_y})], 
the expansion must be to the second order in $\tau v {\bm n} \cdot {\bm q}$ at least. 
Consequently,
\begin{equation}
\langle {\rm e}_x \rangle \sim \tau^2 v^2 (q^2_x - q^2_y), \qquad  
\langle {\rm e}_y \rangle \sim \tau^2 v^2 q_xq_y.
\label{e_xy}
\end{equation}
These terms produce a fourth order correction, $\tau^4v^2 {\bm q}^4$, in the diffusion pole, and, for this reason, can be neglected. 
The average product $\langle {\rm e}_x{\rm e}_y \rangle \sim \tau^4v^2 q^4$ leads to even smaller negligible corrections. 
Next, we note that the main approximation for $\langle {\rm e}^2_{x,y} \rangle$ is
\begin{equation}
\langle {\rm e}^2_{x,y} \rangle \approx {\rm e}^2_{_\|}/2.
\label{e^2_xy}
\end{equation}
Therefore,  Eq. (\ref{Cxyz}) can be approximated as 
\begin{widetext}
\begin{equation}
\left[
\begin{array}{ccc}
\frac{ 1 + {\rm e}^2_z}{2} + 1 - \langle 1 \rangle &
i\langle {\rm e}_z \rangle & 
0\\
- i \langle {\rm e}_z  \rangle & 
\frac{ 1 + {\rm e}^2_z}{2} + 1 - \langle 1 \rangle&
0\\
0 & 
0 &
1 + {\rm e}^2_z - \langle 1-{\rm e}^2_z \rangle
\end{array}
\right]
\left[
\begin{array}{ccc}
 C^{xx} &  C^{xy} & C^{xz} 
\\ 
C^{yx}  &  C^{yy} & C^{yz}
\\ 
C^{zx} & C^{zy} & C^{zz}
\end{array}
\right]
=
\tau
\left[
\begin{array}{ccc}
 1 & 0 & 0 
\\ 
0  & 1 & 0 
\\ 
0  & 0 & 1
\end{array}
\right]. 
\label{Cxyz_1}
\end{equation}
\end{widetext}
It splits into three equations for the required triplet Cooperons:
\begin{equation}
\left[
\begin{array}{cc}
\frac{ 1 + {\rm e}^2_z}{2} + 1 - \langle 1 \rangle &
i{\rm e}_z \langle 1 \rangle \\
- {\rm e}_z i\langle 1 \rangle & 
\frac{ 1 + {\rm e}^2_z}{2} + 1 - \langle 1 \rangle
\end{array}
\right]
\left[
\begin{array}{c}
 C^{xx} \\  C^{yx} 
\end{array}
\right]
=
\tau
\left[
\begin{array}{c}
 1 \\ 0
\end{array}
\right], 
\label{Eq_Cxx}
\end{equation}

\begin{equation}
\left[
\begin{array}{cc}
\frac{ 1 + {\rm e}^2_z}{2} + 1 - \langle 1 \rangle &
i{\rm e}_z \langle 1 \rangle \\
-i{\rm e}_z \langle  1 \rangle & 
\frac{ 1 + {\rm e}^2_z}{2} + 1 - \langle 1 \rangle
\end{array}
\right]
\left[
\begin{array}{c}
 C^{xy} \\  C^{yy} 
\end{array}
\right]
=
\tau
\left[
\begin{array}{c}
 0 \\ 1
\end{array}
\right], 
\label{Eq_Cyy}
\end{equation}

\begin{equation}
[ (1-{\rm e}^2_z)( 1 - \langle 1 \rangle) + 2{\rm e}^2_z ] C^{zz} = \tau. 
\label{Eq_Czz}
\end{equation}
Solving these equations and using Eq. (\ref{Angle_1}) for $\langle 1 \rangle$  
we obtain Cooperon amplitutes (\ref{Cxx}) - (\ref{Czz}).

\end{document}